\documentclass[paper]{JHEP3} 

\JHEPspecialurl{http://jhep.sissa.it/JOURNAL/JHEP3.tar.gz}

\usepackage{epsfig,multicol}

\newcommand\fverb{\setbox\pippobox=\hbox\bgroup\verb}
\newcommand\fverbdo{\egroup\medskip\noindent%
			\fbox{\unhbox\pippobox}\ }
\newcommand\fverbit{\egroup\item[\fbox{\unhbox\pippobox}]}
\newcommand {\beq}{\begin{equation}}
\newcommand {\eeq}{\end{equation}}
\newcommand {\beqa}{\begin{eqnarray}}
\newcommand {\eeqa}{\end{eqnarray}}

\newcommand {\tr}{{\rm tr\,}}

\newcommand {\ee}{\mbox{e}}

\newcommand {\dd}{\mbox{d}}

\newcommand {\del}{\partial}

\newbox\pippobox

\title{Convergence of the Gaussian Expansion Method\\
in Dimensionally Reduced Yang-Mills Integrals}

\author{Jun Nishimura, Toshiyuki Okubo\\
Department of Physics, Nagoya University\\
Furo-cho, Chikusa-ku, Nagoya 464-8602, Japan\\
E-mail: \email{nisimura,okubo@eken.phys.nagoya-u.ac.jp}}
\author{Fumihiko Sugino\\
Service de Physique Th\'{e}orique, C.E.A. Saclay\\
F-91191 Gif-sur-Yvette Cedex, France\\
E-mail: \email{sugino@spht.saclay.cea.fr}}

\preprint{\hepth{0205253}}	

\abstract{
We advocate a method to improve systematically 
the self-consistent harmonic approximation 
(or the Gaussian approximation), which has been 
employed extensively in condensed matter physics 
and statistical mechanics.
Such a method was previously applied to the IIB matrix model,
a conjectured nonperturbative definition of type IIB superstring
theory in ten dimensions. 
Remarkably the dominance of four-dimensional space-time
in the partition function was suggested from calculations
up to the 3rd order. Recently this calculation has been extended to
the 5th order, and the same conclusion has been obtained.
Here we apply this Gaussian expansion method to 
the bosonic version
of the IIB matrix model, where Monte Carlo results are available,
and demonstrate the {\em convergence} of the method 
by explicit calculations up to the 7th order.
More generally we study matrix models 
obtained from dimensional reduction 
of SU($N$) Yang-Mills theory in $D$ dimensions, where the
$D=10$ case corresponds to the bosonic IIB matrix model.
Convergence becomes faster as $D$ increases,
and for $D \gtrsim 10$ it is already achieved at the 3rd order.
}


\begin{document} 


\section{Introduction}
It is well appreciated that perturbation theory has a limited range 
of applicability, although its importance in theoretical physics
can never be overstated.
First of all, perturbative expansions in most cases yield
merely an asymptotic series, which starts to diverge
at some finite order.
There are also situations in which the expansion
parameter is too large to make
perturbative calculations reliable or even meaningful.

Superstring theory provides such a situation where 
nonperturbative effects are considered to be extremely important.
From this point of view, 
the recent proposals for its nonperturbative formulations
should be considered as a substantial progress.
For instance the IIB matrix model \cite{IKKT}
is conjectured to be a nonperturbative definition 
of type IIB superstring theory in ten dimensions,
and it is speculated to explain the dynamical origin of the Standard
Model including the space-time dimensionality, the gauge group
and the number of generations.
Much effort has been made to understand the dynamics of this model.
In particular, Monte Carlo simulations may be as useful
as in lattice gauge theories, and indeed various simplified versions of the
IIB matrix model have been successfully studied \cite{monte,AABHN}.
However, it is not straightforward to extend these studies to
the IIB matrix model, where the {\em complex} fermion determinant
causes the notorious complex-action problem.
In fact it is speculated that the phase of the fermion determinant
plays a crucial role in the dynamical generation of four-dimensional
space-time \cite{NV}.
Recently a new Monte Carlo technique \cite{sign}
is developed to include the effect of the phase.
Preliminary results concerning the space-time dimensionality
are encouraging,
but it remains to be seen whether definite conclusions 
can be reached along this line.
In any case it would be certainly desirable to
develop an alternative method which enables a nonperturbative access to
the dynamics of the model.

In this Letter, we are going to advocate 
a method to systematically improve
the self-consistent harmonic approximation 
(or the Gaussian approximation), which has been widely used in condensed
matter physics. 
(See Ref.\ \cite{Iucci:2001ym} and references therein.)
Indeed such a method
was applied to the IIB matrix model up to the 3rd order 
in Ref.\ \cite{Nishimura:2001sx}.
This provided the first analytical evidence for the dynamical generation of 
{\em 4d space-time}, 
which is both surprising and encouraging 
since the model describes superstrings in {\em 10 dimensions}.
Stability of this conclusion against higher order corrections
has been confirmed recently by explicit 5th order calculations \cite{KKKMS}.
There the method is also interpreted as an improved Taylor expansion.
The Gaussian approximation (the leading order of the Gaussian
expansion) has been applied to random matrix models earlier in 
Ref.\ \cite{EL}.
Its application to supersymmetric matrix quantum mechanics
(including the Matrix Theory)
was advocated by Kabat and Lifschytz \cite{Kabat:2000hp},
and Refs.\ \cite{blackholes} succeeded in
revealing interesting blackhole thermodynamics
even at the leading order.

Here we apply the method to 
the bosonic version of the IIB matrix model.
More generally we consider the model \cite{Krauth:1998yu} obtained
from dimensional reduction 
of SU($N$) Yang-Mills theory in $D$ dimensions,
where the $D=10$ case corresponds to the bosonic IIB matrix model.
The model can also be regarded as a Hermitian matrix version of the
Eguchi-Kawai model \cite{EK}, which was proposed as an equivalent
description of large-$N$ gauge theory.
This equivalence in the case of the present model has been
investigated recently \cite{ABN}.
Hence we expect that our results would also have some implications
to field theoretical applications of the method.

In the present bosonic model,	
we perform calculations up to 
the 7th order, which allows us to
convincingly demonstrate the convergence of the method.
Although the model is much simpler than the IIB matrix model,
it is still nontrivial, and in particular 
it has not been solved exactly so far.
Also the conventional perturbative approach cannot be applied
as in the IIB matrix model,
since the action does not contain a quadratic term.
However, standard Monte Carlo simulation is applicable,
and there are some results \cite{HNT} that can be compared with ours.
As another advantage of studying this model,
we can perform analytic calculations leaving $D$ as a free parameter.
We observe that the convergence is fast in particular for large $D$.
For $D \gtrsim 10$, for instance, a reasonable convergence is already
achieved at the 3rd order, which supports the validity
of the aforementioned calculations in the IIB matrix model.

We would like to point out that actually 
the Gaussian expansion method
has a long history \cite{conv,Stevenson:1981vj},
although its application to matrix models is new.
Originally it was developed to obtain the energy spectrum
in quantum mechanical systems.
The most striking aspect of the method
is that the results {\em converge},
although the expansion is {\em not} based on any small parameter
\footnote{There are rigorous proofs of convergence 
in certain examples \cite{exact_conv}. 
Very recently Ref.\ \cite{largest}
has shown that complex solutions to
the $d(\mbox{result})/d(\mbox{unphysical parameter})=0$ 
equation (in fact, the solution with the largest imaginary part) 
works extremely well.
}.
It is also interesting that the results contain fully nonperturbative
effects, although the actual calculations are
nothing but the familar ones in standard perturbation theory.
In this approach one typically encounters a situation in which
the approximants depend on free parameters.
In fact the power of the method comes from the flexibility to adjust
these parameters depending on the order of the expansion.
Common wisdom suggests to determine them by 
`the principle of minimum sensitivity' \cite{Stevenson:1981vj}.
This is usually achieved by solving the `self-consistency equation'.
However, we will encounter some problems with this strategy.
%
%
Here we propose a novel prescription
based on histograms,
with which one can naturally 
obtain the `best approximation' at each order
without such problems.
We hope that our histogram prescription is useful
in other applications as well.
%
%
%

%

\section{Dimensionally reduced Yang-Mills integrals}

The model we study in this work
is defined by the partition function
\beqa
Z &=& \int \dd A \, \ee ^{-S} \ ,
\label{bosonicZ} \\
S &=& - \frac{1}{4}N\beta \, \sum_{\mu , \nu}
\tr [A_\mu , A_\nu] ^2  \ ,
\label{bosonicSA}
\eeqa
where $A_\mu$ ($\mu = 1, \cdots , D$) 
are $N\times N$ Hermitian matrices.
The integration measure $\dd A$ is defined by
$ \dd A = \prod_{a=1}^{N^2-1} \prod_{\mu = 1}^{D}
\frac{d A_\mu ^a}{\sqrt{2 \pi}} $,
where $A_\mu^a$ are the coefficients in
the expansion
$A_\mu = 
\sum_{a}
A_\mu ^a \, T^a $
with respect to the SU($N$) generators 
$T^a$ normalized as $\tr (T^a T^b) = \frac{1}{2} \delta ^{ab}$.
The model can be regarded as the zero-volume limit
of SU($N$) Yang-Mills theory in $D$ dimensions.
As a result of this limit,
we can actually absorb $\beta$ in (\ref{bosonicSA})
by rescaling $A_\mu \mapsto \beta^{-1/4}
A_\mu$.
Thus $\beta$
is not so much a coupling constant as a scale parameter,
which we set to $\beta = 1$ without loss of generality.
The partition function was conjectured \cite{Krauth:1998yu}
and proved \cite{Austing:2001bd} to be finite 
for $N > D/(D-2)$.
Note in particular that the model is ill-defined for $D \le 2$.
A systematic $1/D$ expansion was formulated in \cite{HNT}.
In particular the absence of SO($D$) breaking was shown 
to all orders of the $1/D$ expansion
and this conclusion was also confirmed by 
Monte Carlo simulations \cite{HNT}
for various $D=3,4,6,\cdots ,20$.
The Gaussian expansion method is also capable of
addressing such an issue \cite{Nishimura:2001sx},
but here we assume the absence of the SSB
in order to focus on the convergence of the method itself.
In Ref.\ \cite{Oda:2001im}
Polyakov line and Wilson loop were calculated
by the (next-)leading Gaussian approximation,
and Monte Carlo results \cite{AABHN} were reproduced qualitatively.
An equally successful result was obtained 
in the supersymmetric case \cite{Sugino:2001fn}.

\section{The Gaussian expansion method}

The Gaussian action which has both SU($N$) and SO($D$) symmetries
can be written as
\beq
S_0 = \frac{N}{v} \frac{\sqrt{D}}{2} \sum _{\mu = 1} ^{D}  
\tr \left( A_\mu \right)^2 \ , 
\label{S0def}
\eeq
where $v$ is a positive real parameter.
(The constant factor $\frac{\sqrt{D}}{2}$ is introduced for convenience.) 
Then we rewrite the partition function (\ref{bosonicZ}) as
\beqa
Z &=& Z_0  \, \langle  \ee ^{- (S-S_0)} \rangle _0 \ , 
\label{factori}
\\
Z_0 &=& \int \dd A \, \ee ^{-S_0}
 \ ,
\label{defZ0}
\eeqa
where $\langle \ \cdot \ \rangle_0$ is a VEV with respect to the
partition function $Z_0$.
{}From (\ref{factori}) it follows that the free energy $F = - \ln Z$
can be expanded as
\beqa
\label{free_expand}
F &=& \sum_{k=0}^{\infty} F_k  ~~~;~~~
F_0 \equiv - \ln Z_0 \ , \\
F_k &\equiv&  -  \frac{(-1)^k}{k!} \langle (S - S_0)^k \rangle_{{\rm C},0} 
~~~~~~~~(\mbox{for}~k\ge1) \ ,
\eeqa
where the suffix `C' in $\langle \ \cdot \ \rangle _{{\rm C} , 0}$
means that the connected part is taken.
Each term $F_k$ in the above expansion 
can be calculated conveniently by using Feynman diagrams.
In actual calculations we have to truncate 
the infinite series (\ref{free_expand}) at some finite order.
%
In Ref.\  \cite{KKKMS}, Schwinger-Dyson equations were used 
to reduce the number of diagrams considerably. 
Using this technique, we were able to proceed up to the 7th
order in the present model with reasonable efforts
({\em e.g.}, we evaluated 21 eight-loop diagrams.).


\FIGURE{\epsfig{file=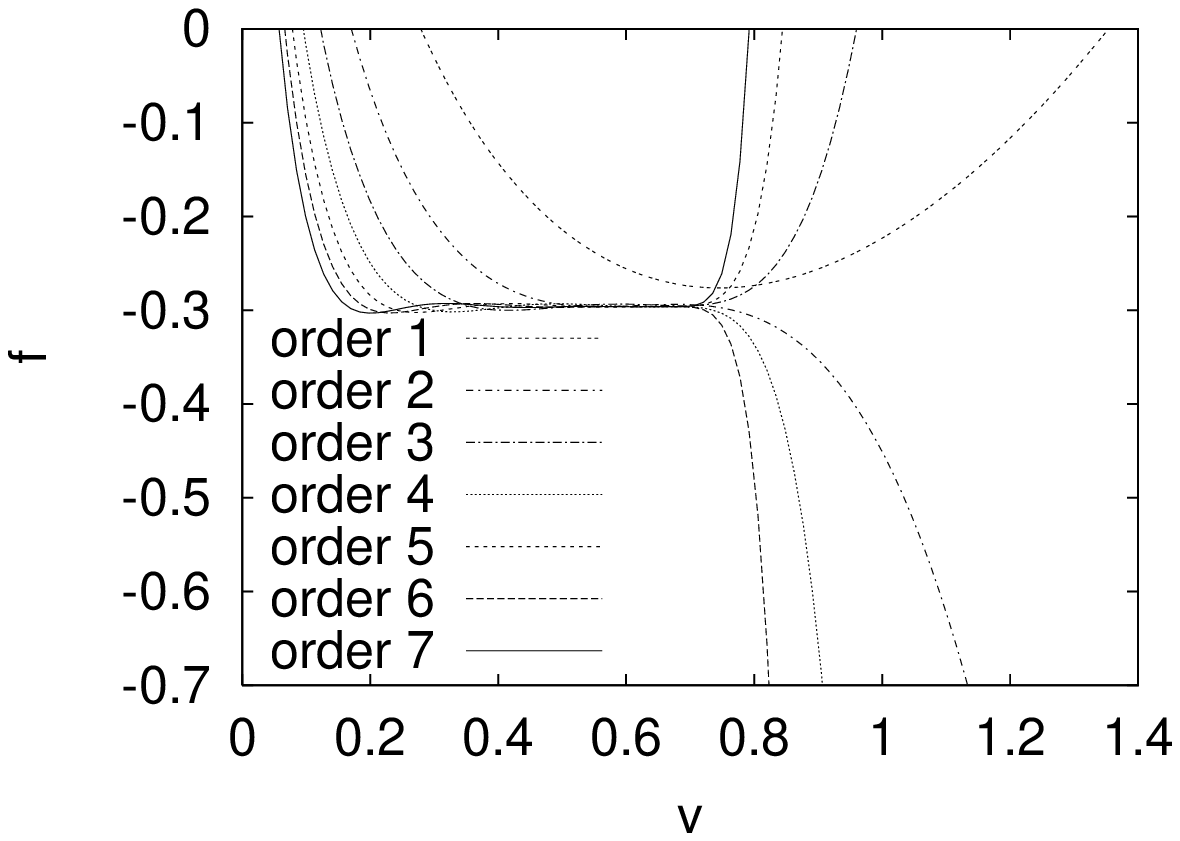,width=10cm} 
    \caption{The truncated free energy density $f$ 
is plotted as a function of 
$v$
for $D = 10$.
Each curve corresponds to order 1, 2, $\cdots$, 7.
Formation of a plateau is clearly seen.
}
    \label{fig:freeE_10d}
}

Once we truncate the expansion (\ref{free_expand}),
the calculated free energy
depends on the parameter $v$ introduced in the Gaussian action (\ref{S0def}).
In the large-$N$ limit 
\footnote{This simply amounts to considering only the planar diagrams.
Although it reduces our effort considerably,
the method itself is applicable to arbitrary finite $N$.
}
we compute the `free energy density' defined by
\beq
f = 
\lim _{N\rightarrow \infty} \left\{
\frac{1}{D(N^2 -1)}F 
- \frac{1}{2} \ln \left( \sqrt{\frac{D}{2}}N \right) 
\right\} 
\ .
\eeq
The second term is subtracted in order to make the quantity finite.
(In particular $f = -  1/4$ for $D=\infty$.)
%
In Fig.\ \ref{fig:freeE_10d} we plot the result at each order
as a function of $v$ for $D=10$.
Formation of a plateau is clearly seen.
We may understand this phenomenon as a reflection of the fact that
the formal expansion (\ref{free_expand}), if {\em not}
truncated, should not depend on $v$.
According to this interpretation, the height of the plateau observed
for the truncated free energy is expected to provide a good approximation.
This is the philosophy behind `the principle of minimum sensitivity',
which is known in a more general context \cite{Stevenson:1981vj}.

\FIGURE{\epsfig{file=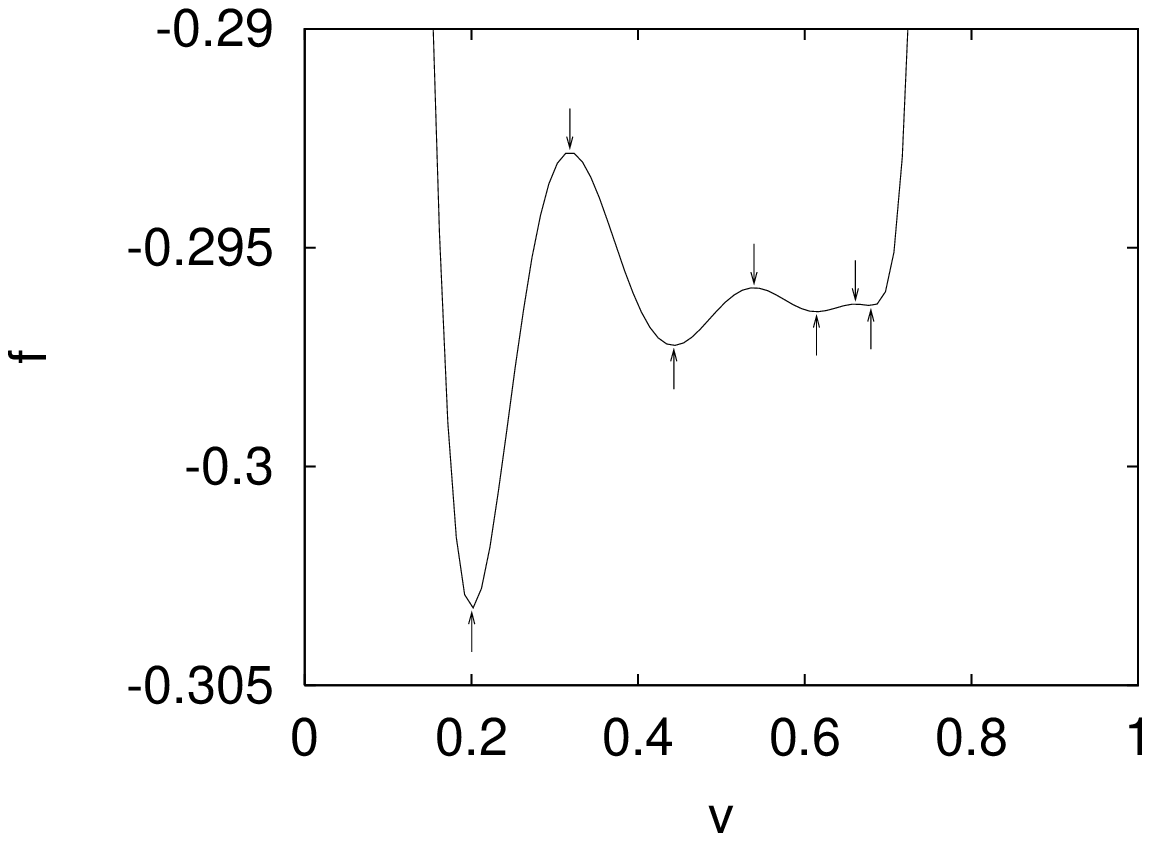,width=10cm} 
    \caption{
The zoom up of Fig.\ \ref{fig:freeE_10d} for order 7.
The arrows indicate the position of the extrema,
each of which corresponds to a solution of the 
`self-consistency equation'.
}
    \label{fig:zoomup_freeE_histo_10d}
}


\TABLE[pos]{
\begin{tabular}{|c||c|c|c|c|c|c|c|}
\hline
$D$ 
 & \multicolumn{7}{c|}{{\em order} ($n$)} \\  \cline{2-8}
&  1 & 2 & 3 & 4 &5 & 6 & 7\\
\hline
3 &  1 &  0 &  1 &  0 &  1 &  0 &  1 \\
4 &  1 &  0 &  1 &  0 &  1 &  0 &  1 \\
5 &  1 &  0 &  1 &  0 &  1 &  2 &  5 \\
6 &  1 &  0 &  3 &  2 &  5 &  4 &  7 \\
7 &  1 &  2 &  3 &  4 &  5 &  6 &  7 \\
$\vdots$ &  1 &  2 &  3 &  4 &  5 &  6 &  7 \\
$\infty$ &  1 &  2 &  3 &  4 &  5 &  6 &  7 \\
\hline
\end{tabular}
\caption{The number of solutions to the `self-consistency 
equation' (\ref{self-consistency_eq}) for each $D$. 
}
\label{table:no_of_solutions}
}

Next we need to specify a prescription to obtain a concrete value 
which approximates the free energy at each order.
In Refs.\ \cite{Nishimura:2001sx,KKKMS} the free parameters in the Gaussian
action were determined such that the truncated free energy becomes
stationary.
This amounts to solving the
`self-consistency equation'
\beq
\frac{\del}{\del v} \left( \sum_{k=0}^{n} F_k \right) = 0  \ .
\label{self-consistency_eq}
\eeq
In Table \ref{table:no_of_solutions}
we list the number of solutions to the 
`self-consistency equation' for each $D$, which already reveals 
some problems.
For example, for $D=10$ we find that the number of solutions
increases as we go to higher orders, and we have to decide which one 
to choose.
A similar ambiguity was encountered in the 
IIB matrix model \cite{Nishimura:2001sx,KKKMS},
where the solution that gives the smallest free energy has been 
chosen.
While this prescription seems reasonable at relatively small orders,
we have problems at higher orders, in particular when we are discussing
the convergence.
In Fig.\ \ref{fig:zoomup_freeE_histo_10d}
we zoom up the plateau seen at the order 7 in Fig.\ \ref{fig:freeE_10d}.
We observe some oscillations,
whose amplitude becomes larger towards the left edge of the plateau. 
If we used the prescription described above in the present situation,
we would clearly underestimate the height of the plateau.
This will be even more problematic at higher orders
since the dip at the left edge seems to become deeper 
as the order increases.
It may also happen in general that the truncated free energy
acquires a global minimum which has nothing to do with the
plateau formation.

%
From Table \ref{table:no_of_solutions}
one also finds that
there are cases where the `self-consistency equation' 
has no solutions.
(This was encountered in the study of the IIB matrix model
at the orders 2 and 4 \cite{Nishimura:2001sx,KKKMS}.)
For instance there is no solution at the orders 2,4,6 for $D=4$.
As we show in the Appendix, a well-defined plateau is developing
even in this case.
The question is how to extract the height of the plateau in such
cases.

\section{The histogram prescription}
\label{histogram}

All these problems arise from trying to {\em determine} the free
parameters by local information such as
`self-consistency equations'.
In fact it is more natural to extract the height of the plateau
{\em directly}.
%
As a first step, let us make a histogram of the truncated free energy 
calculated at points which
are uniformly distributed in the parameter space of the Gaussian
action.
Fig.\ \ref{fig:freeE_histo_10d_sbin} shows the histogram
for $D=10$ at order 7.
Whenever we make histograms in what follows,
the range of $v$ is restricted to $0.01 \le v \le 1.4$, and the
vertical axis is normalized such that the integration gives unity.
In Fig.\ \ref{fig:freeE_histo_10d_sbin}
the bin size is chosen to be very small,
and as a result we obtain many peaks, 
each of which corresponds to a solution of the `self-consistency equation'
(Actually the highest peak contains the two solutions which lie
at the right edge of the plateau in 
Fig.\ \ref{fig:zoomup_freeE_histo_10d}.).
Hence we are back in the previous situation, and we cannot reasonably
extract the height of the plateau.
In order to identify the plateau, we clearly need some kind of 
{\em coarse graining}.
Let us note here that
the bin size of the histogram represents the precision at which we
search for the plateau.
Therefore, we increase the bin size gradually until a single
peak becomes `dominant'. 
As a criterion of the dominance, we require
the highest bin to be more than twice
as high as the second highest one.
We further require the dominance to persist for larger bin size.
The resulting histogram is shown in 
Fig.\ \ref{fig:freeE_histo_10d_lbin}.

\FIGURE{\epsfig{file=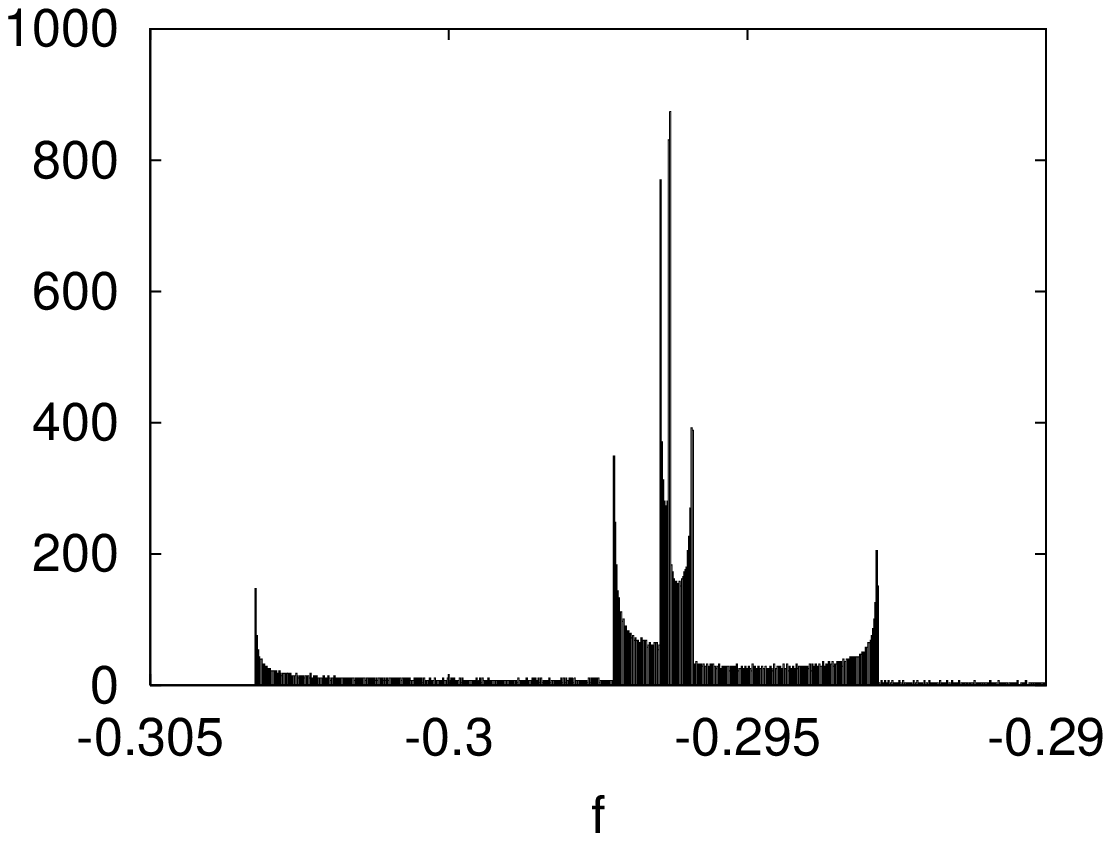,width=10cm} 
    \caption{
The histogram of the truncated 
free energy density $f$ for $D=10$ at order 7.
The bin size is chosen to be very small (0.00002).
}
    \label{fig:freeE_histo_10d_sbin}
}

\FIGURE{\epsfig{file=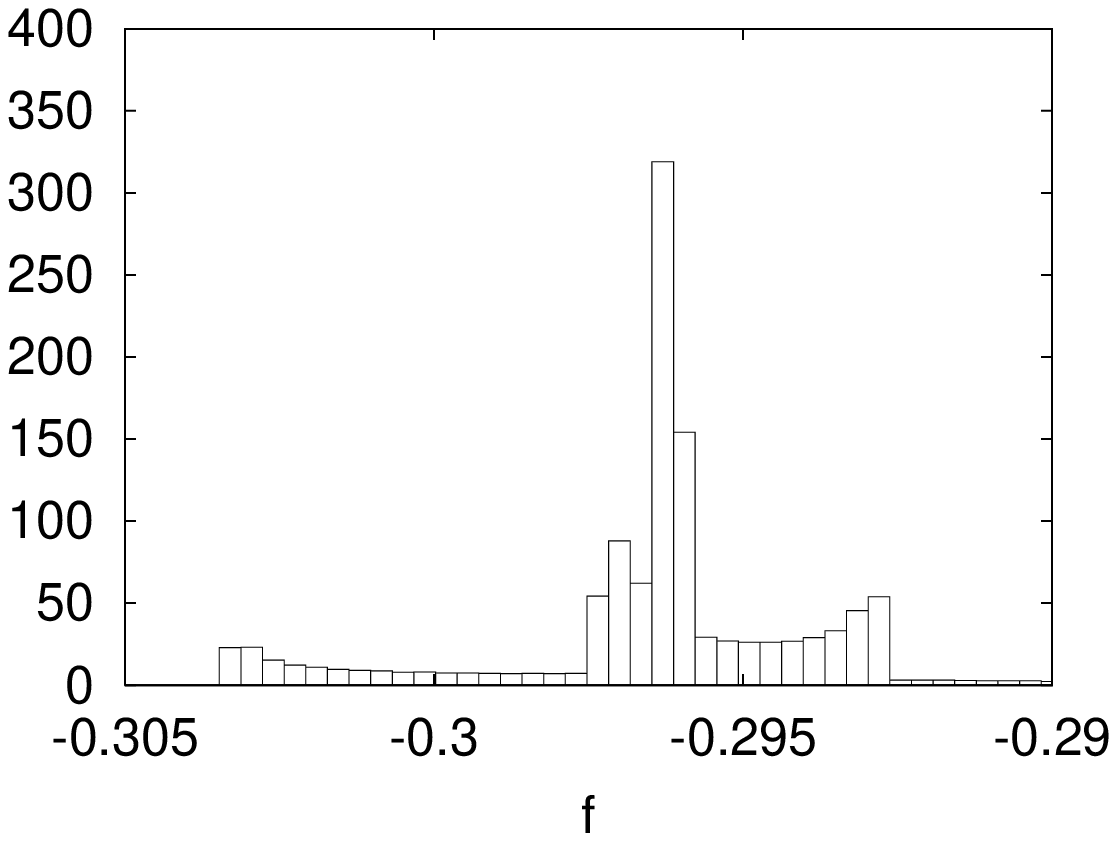,width=10cm} 
    \caption{
The histogram of the truncated 
free energy density $f$ for $D=10$ at order 7.
The bin size (0.00035) 
is chosen such that the highest bin is more than
twice as high as the second highest one.
}
    \label{fig:freeE_histo_10d_lbin}
}

\FIGURE{\epsfig{file=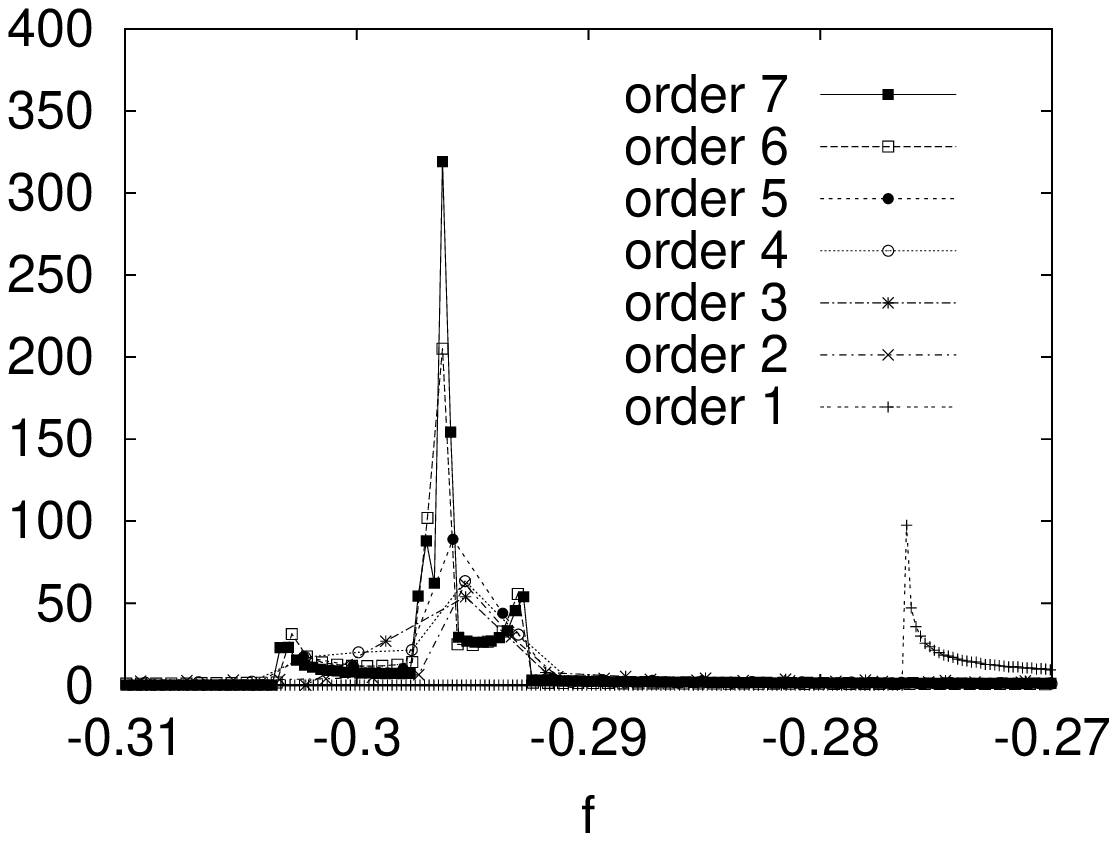,width=10cm} 
    \caption{
The histogram of the truncated free energy density $f$ for $D = 10$ 
at order 1, 2, $\cdots$, 7. 
The bin size is chosen carefully as described in the text at each
order.
The growth of the peak is clearly observed.
}
    \label{fig:freeE_histo_10d}
}



Fig.\ \ref{fig:freeE_histo_10d} shows the histograms obtained in this
way at each order for $D=10$.
The peak becomes sharper and sharper as we go to 
higher order. 
Thus the histogram serves as a clear indicator of the plateau formation.
Moreover, now we can obtain the `best approximation' at each order
from the position of the peak,
while the bin size (chosen by the above criterion) 
roughly represents the theoretical uncertainty of the method.
In the Appendix, we present a similar analysis for $D=4$.
In this case the `self-consistency equation' has no solutions
at orders 2,4,6 as mentioned above.
However, the histogram prescription is as successful as in $D=10$.

In Fig.\ \ref{fig:freeE_order}, 
we plot the approximated value of the free energy for various $D$
obtained at each order by our histogram prescription.
The `error bars' represent a rough estimate for the theoretical 
uncertainty explained above 
\footnote{We did not put error bars to the order 1 results.
This is because our criterion for the dominance of a single bin
is satisfied even for an infinitesimal bin size.
Thus at the order 1, the histogram prescription does nothing more than
just solving the `self-consistency equation'.
}.
(Remark: Unlike usual error bars, they do not represent an estimate 
for possible discrepancies from the correct value.
For the latter, one should also consider the changes
of the results as the order is increased.)
We observe a clear convergence for each $D$.
The convergence becomes fast in particular for large $D$.

\FIGURE{\epsfig{file=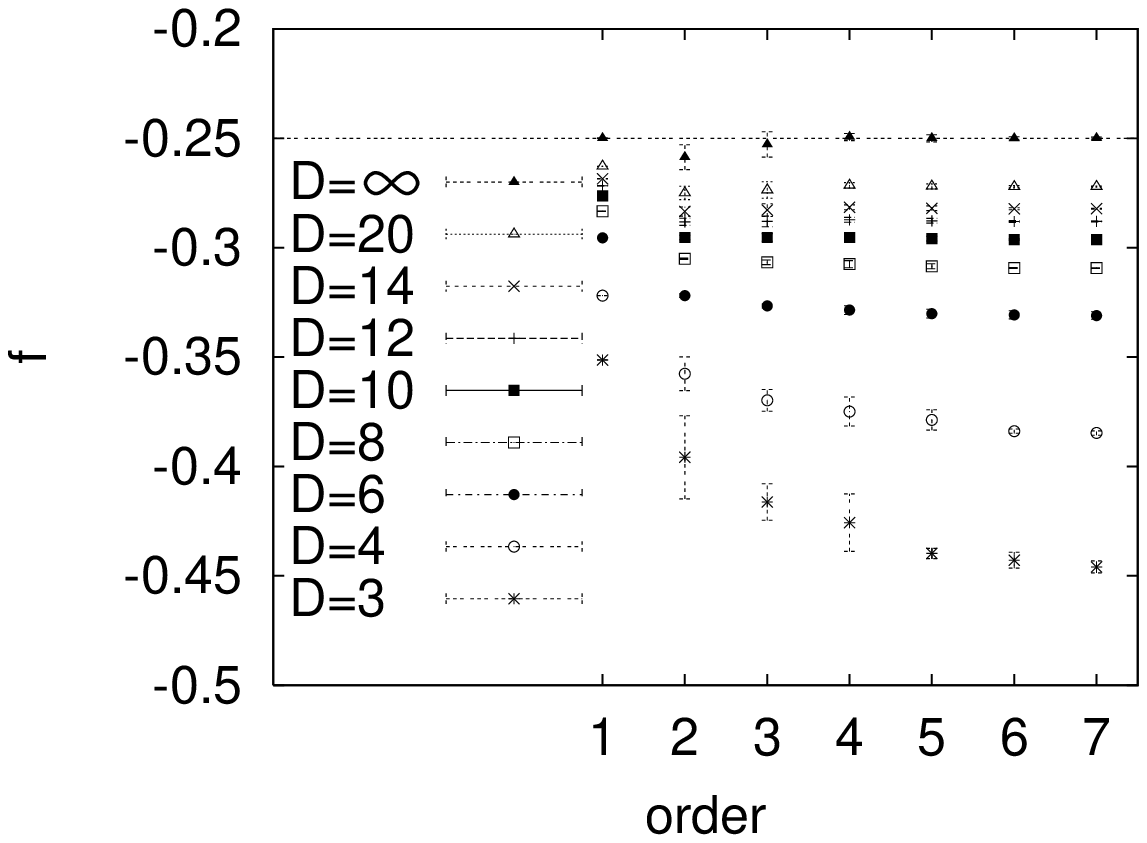,width=10cm} 
    \caption{
Free energy density $f$ obtained by
the histogram prescription at order 1, 2, $\cdots$, 7
for various $D$. The horizontal line represents the exact result
$f=-1/4$ for $D=\infty$.
}
    \label{fig:freeE_order}
}

\section{Observables}
\label{results}

Similarly to the free energy (\ref{free_expand}),
the expectation value of
an operator ${\cal O}$ is expanded as
\beqa
\label{operator_inf}
\langle {\cal O} \rangle &=& \sum_{k=0}^{\infty}
O_k  ~~~;~~~
O_0 \equiv \langle {\cal O} \rangle_0 \ ,
\\
O_k
& \equiv &  
 \frac{(-1)^k}{k!} \, 
\Bigl\langle (S - S_0 )^{k} {\cal O}
\Bigr \rangle_{{\rm C},0}
~~~~~(\mbox{for}~k\ge1) \ .
\label{operator_k}
\eeqa
As a fundamental object in this model \cite{HNT},
we consider
\beq
{\cal O} = \frac{1}{N} \sum _{\mu = 1} ^{D}  
\tr \left( A_\mu \right) ^2 \times \sqrt{\frac{2}{D}} \ .
\eeq
The normalization is chosen such that
$\langle {\cal O} \rangle = 1$ for $D=\infty$ in the 
large-$N$ limit \cite{HNT}.
We calculate $\langle {\cal O} \rangle$ by the series expansion
(\ref{operator_inf}) truncated at some order.
In Fig.\ \ref{fig:extent_10d} we plot the result at each order
as a function of $v$ for $D=10$.

In Refs.\ \cite{Nishimura:2001sx,KKKMS},
the truncated expectation value was evaluated at the same value of $v$
as the one used to evaluate the free energy.
Note, however, that the region of the parameter $v$ where
the plateau develops for the expectation value
is slightly shifted from the corresponding region 
for the free energy.
It is therefore safer to extract an approximated value for the observable
by repeating the same procedure as we did for the free energy.

Fig.\ \ref{fig:extent_histo_10d} shows the results of the histogram
prescription at each order.
The growth of the peak is clearly observed, which confirms the
plateau formation.
Fig.\ \ref{fig:extent_order} summarizes the final results 
for the observables, where the `error bars' again represent
the theoretical uncertainty.
Comparison with the Monte Carlo data \cite{HNT}
confirms that our results do converge to the correct values.
For $D\gtrsim 10$ in particular, 
the convergence is already achieved at order 3.
%

It was previously noted that 
the leading Gaussian approximation becomes
`exact' at $D=\infty$ \cite{Oda:2001im}.
The precise statement is that
if we evaluate the free energy density and the observable 
(truncated at the 1st order)
at the value of $v$ which solves the 1st order `self-consistency
equation' {\em for the free energy},
they agree with the exact results at
$D = \infty$.
Although this can be naturally understood 
from the viewpoint of the $1/D$ expansion \cite{HNT},
it looks rather accidental 
in the present framework.
%
%
If we evaluate the truncated observable 
at $v$ which solves
the `self-consistency equation' {\em for the observable},
we obtain 1.08866 instead of 1.
Also there is no plateau yet at order 1.
Figs.\ \ref{fig:freeE_order} and \ref{fig:extent_order}
show that the convergence of the method as it stands
is achieved at order 3 (but not yet at order 1)
even for $D=\infty$.

\FIGURE{\epsfig{file=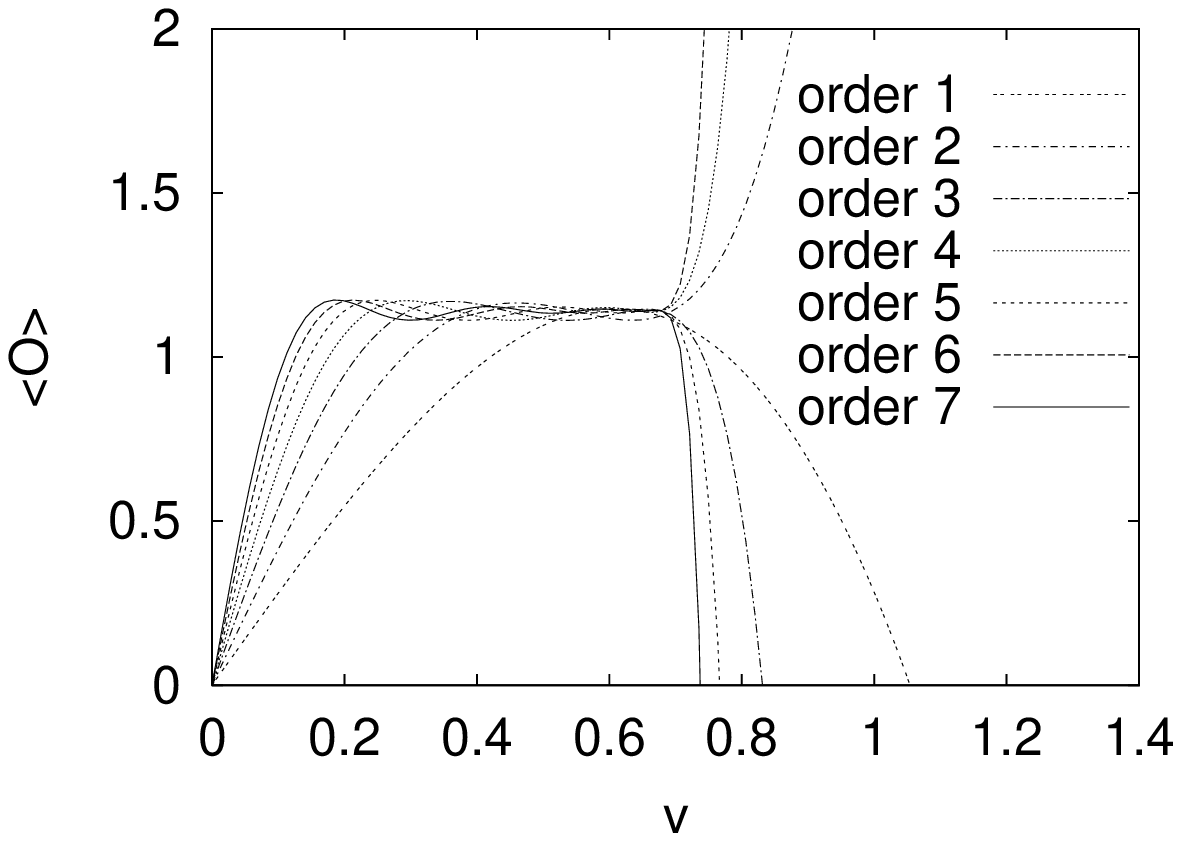,width=10cm} 
    \caption{
The truncated expectation value $\langle {\cal O} \rangle$
is plotted as a function of $v$
for $D = 10$.
Each curve corresponds to order 1, 2, $\cdots$, 7.
Formation of a plateau is clearly seen.
}
    \label{fig:extent_10d}
}

\FIGURE{\epsfig{file=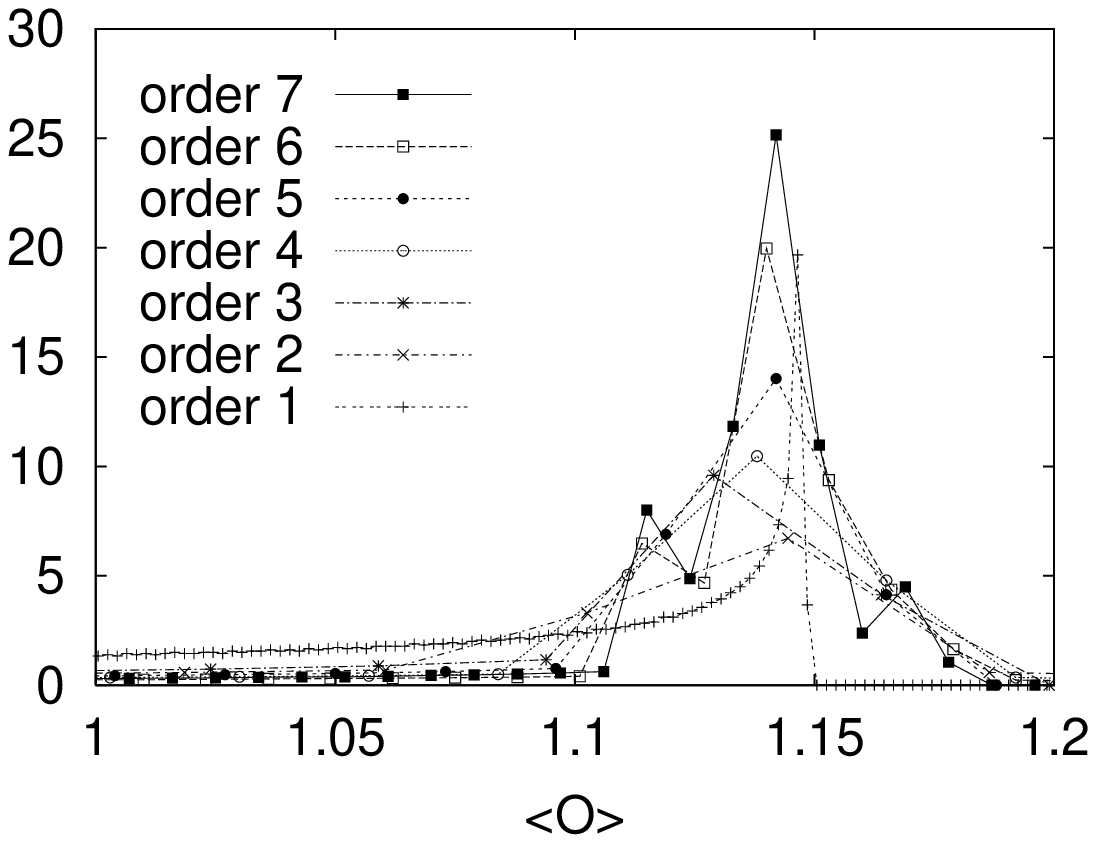,width=10cm} 
    \caption{
The histogram of 
the truncated expectation value $\langle {\cal O} \rangle$
for $D = 10$ 
at order 1, 2, $\cdots$, 7. 
The bin size is chosen carefully as described in the text at each
order.
The growth of the peak is clearly observed.
}
    \label{fig:extent_histo_10d}
}

\FIGURE{\epsfig{file=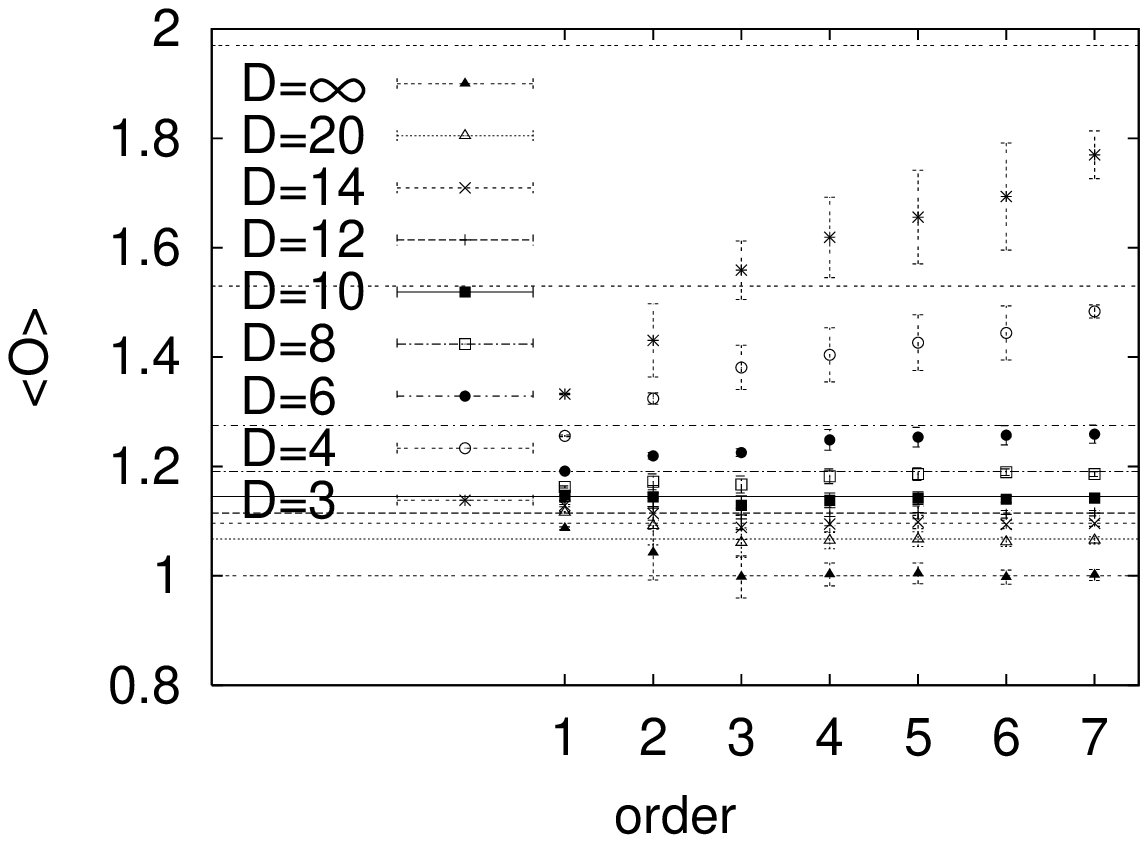,width=10cm} 
    \caption{
The expectation value $\langle {\cal O} \rangle$
obtained by the histogram prescription at order 1, 2, $\cdots$, 7
for various $D$. 
The horizontal lines represent the Monte Carlo results
for each $D<\infty$ and the exact result $\langle {\cal O}\rangle = 1$
for $D=\infty$.
}
    \label{fig:extent_order}
}


In this regard,
it should be emphasized that the Gaussian expansion is {\em different}
from the $1/D$ expansion.
Each term in the expansions for the free energy density
and the observable
is {\em of order 1} with respect to $1/D$ for generic $v$.
Still the height of the plateau converges due to nontrivial
cancellations among diagrams.
%
Note also that the Gaussian expansion can be applied to the 
supersymmetric case (including the IIB matrix model),
while the $1/D$ expansion cannot be formulated there.
%




\section{Summary and Discussions}
\label{summary}

The most important conclusion of our study is that 
the Gaussian expansion method converges in the dimensionally
reduced Yang-Mills integrals.
In particular the convergence is achieved already at order 3
for $D=10$, which corresponds to the bosonic IIB matrix model.
This suggests the particular usefulness of the method in studying
the IIB matrix model.
In demonstrating the convergence we encountered problems associated
with the commonly used prescription based on `self-consistency
equations'.
We solved these problems by extracting the height of the plateau
directly using a novel prescription based on histograms.

The histogram prescription is expected to be 
particularly useful in the case where the parameter space 
of the Gaussian action becomes multi-dimensional.
If there are more than two parameters, it is already difficult to
visualize the plateau formation as we did in 
Figs.\ \ref{fig:freeE_10d}, \ref{fig:extent_10d} and
\ref{fig:freeE_4d}.
When we study the SSB of SO(10) symmetry
in the IIB matrix model, we have to introduce
10 real and 120 complex parameters \cite{Nishimura:2001sx}.
Here we will need some kind of important sampling technique
like Monte Carlo simulations in order to explore such a huge parameter 
space.
In the previous works \cite{Nishimura:2001sx,KKKMS}
based on `self-consistency equations', 
there were problems like the ones we saw in this work.
At order 2 and 4 there were no solution to the 
`self-consistency equations',
and in the other cases the solution was not unique. 
We are currently reinvestigating the IIB matrix model
by using the histogram prescription with the above extension
\cite{prep}.

Finally we would like to emphasize that
the dimensionally reduced model
we studied in this paper is a system
of {\em infinitely many degrees of freedom},
since we have taken the size of the matrices $A_\mu$ to be 
infinite.
In this regard, let us recall that the model
has also connections to SU($N$) gauge theories in $D$-dimensional 
space time in the large-$N$ limit \cite{EK}.
The dimensional reduction amounts to reducing the
base space of the SU($N$) gauge theories to a point,
but the original space-time degrees of freedom are still
somehow encoded in the internal degrees of freedom.
This is possible precisely because there are infinitely many
internal degrees of freedom after taking the large-$N$ limit. 
From this point of view 
the convergence of the Gaussian expansion method in the present model
is interesting since the examples for which
the convergence is shown in the literature
have been restricted to simple quantum-mechanical systems.
Although higher order calculations in 
field-theoretical applications would be technically more involved 
due to the necessity of the renormalization procedure 
(See, however, Ref.\ \cite{Stancu:sk}),
we hope that our results open up a new perspective
in that direction as well.



\acknowledgments
We would like to thank the authors of Ref.\ \cite{KKKMS}
for stimulating discussions.
The work of J.N.\ is supported in part by Grant-in-Aid for 
Scientific Research (No.\ 14740163) from 
the Ministry of Education, Culture, Sports, Science and Technology.


\appendix

\section{The situation in $D=4$ : plateau without
extrema}\label{sec:A}


In this Appendix, we present our results for $D=4$.
As we see from Fig.\ \ref{fig:freeE_4d},
a well-defined plateau is clearly developing.
However, if we zoom up the curves for order 2,4,6
(Fig.\ \ref{fig:zoomup_freeE_histo_4d}),
we find that there are no extrema,
in agreement with the absence of 
solutions to the `self-consistency equation' 
(See Table \ref{table:no_of_solutions}).
We will see that the histogram prescription works even in this case.

Figs.\ \ref{fig:freeE_histo_4d_sbin} and
\ref{fig:freeE_histo_4d_lbin}
show the histograms at order 6 before and after `coarse-graining'.
As we increase the bin size gradually, a single 
bin becomes dominant 
(See Section \ref{histogram} for the precise criterion.).
Thus the histogram prescription allows us to obtain
explicit results at each order including orders 2,4,6, where
the `self-consistency equation' has no solutions.
Fig.\ \ref{fig:freeE_histo_4d} shows the results at
orders $1,\cdots ,7$.
The growth of the peak is clearly observed, which 
confirms the plateau formation.
Moreover, the results obtained in this way show
a clear convergence
as we have seen in Fig.\ \ref{fig:freeE_order}.


\FIGURE{\epsfig{file=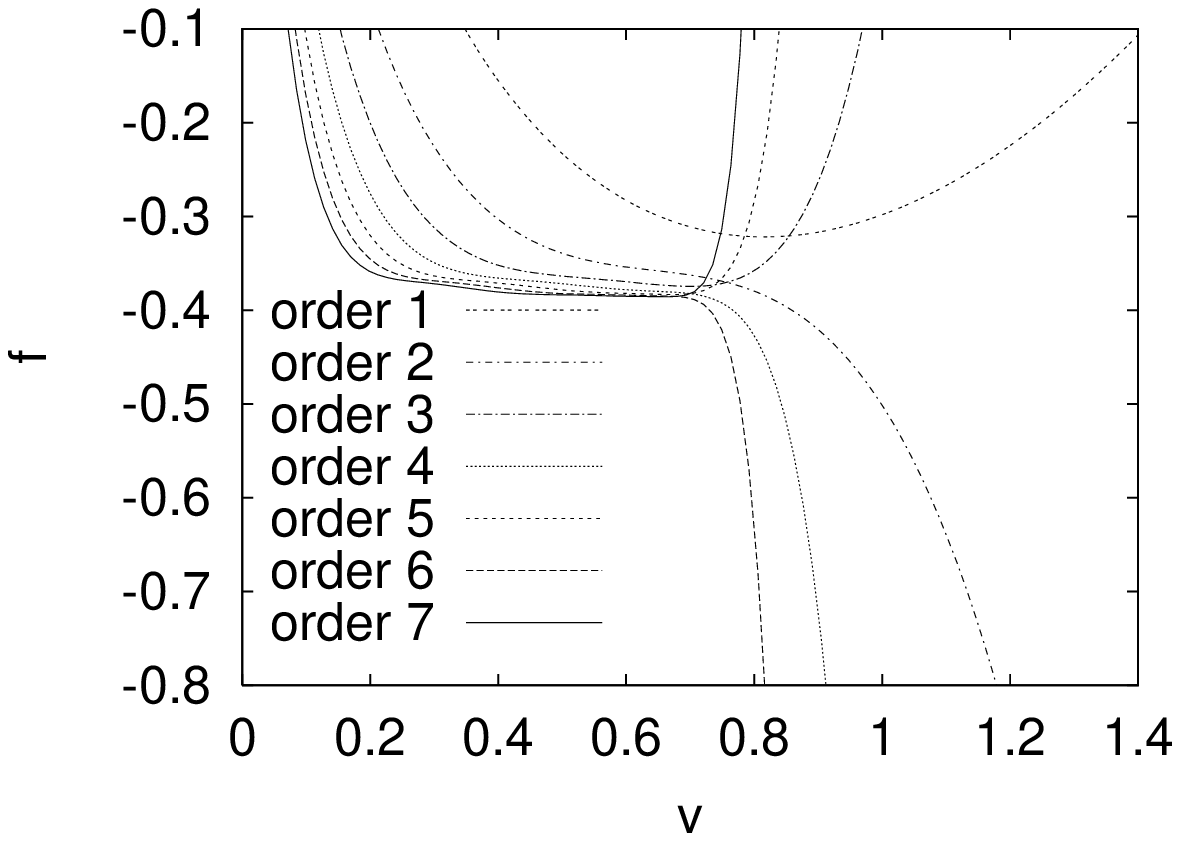,width=10cm} 
    \caption{The truncated 
free energy density $f$ is plotted as a function of 
$v$
for $D = 4$.
Each curve corresponds to order 1, 2, $\cdots$, 7.
Formation of a plateau is clearly seen.
}
    \label{fig:freeE_4d}
}

\FIGURE{\epsfig{file=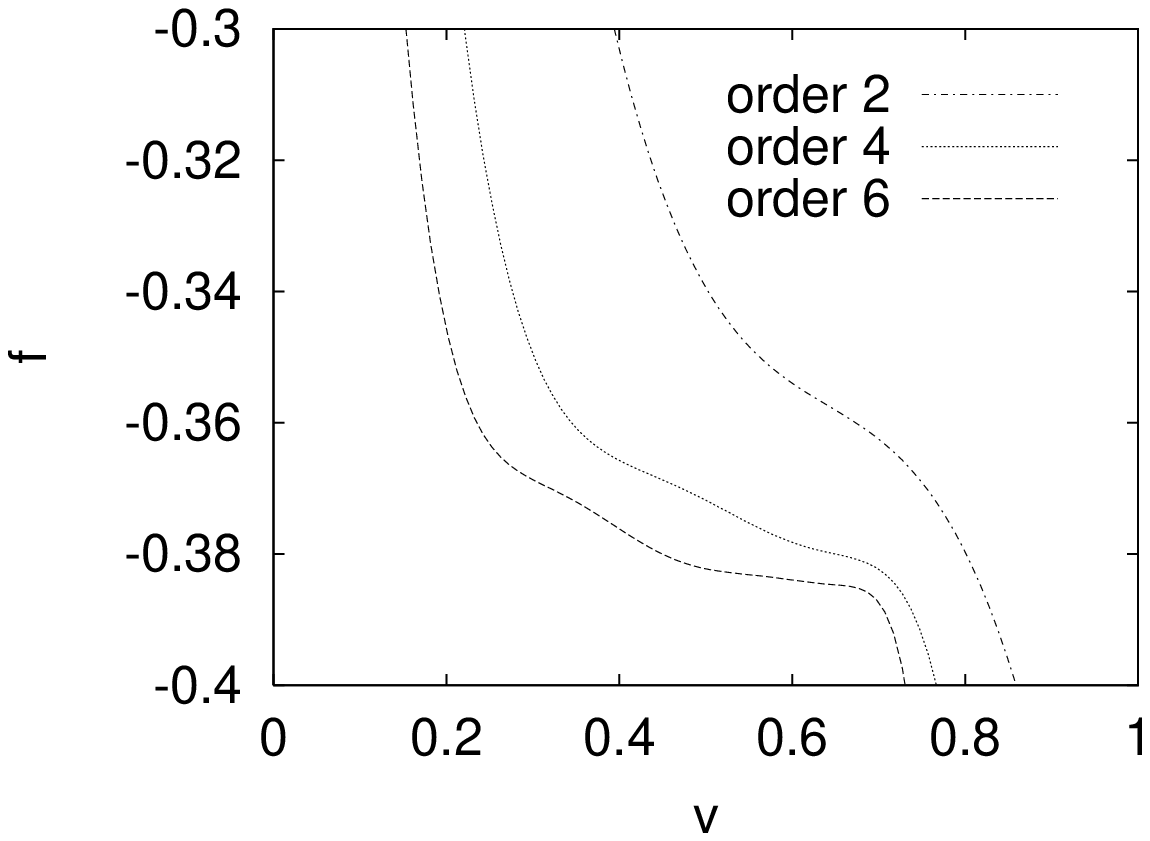,width=10cm} 
    \caption{
The zoom up of Fig.\ \ref{fig:freeE_4d} for order 2,4,6.
There is no extremum point in the curves.
}
    \label{fig:zoomup_freeE_histo_4d}
}

\FIGURE{\epsfig{file=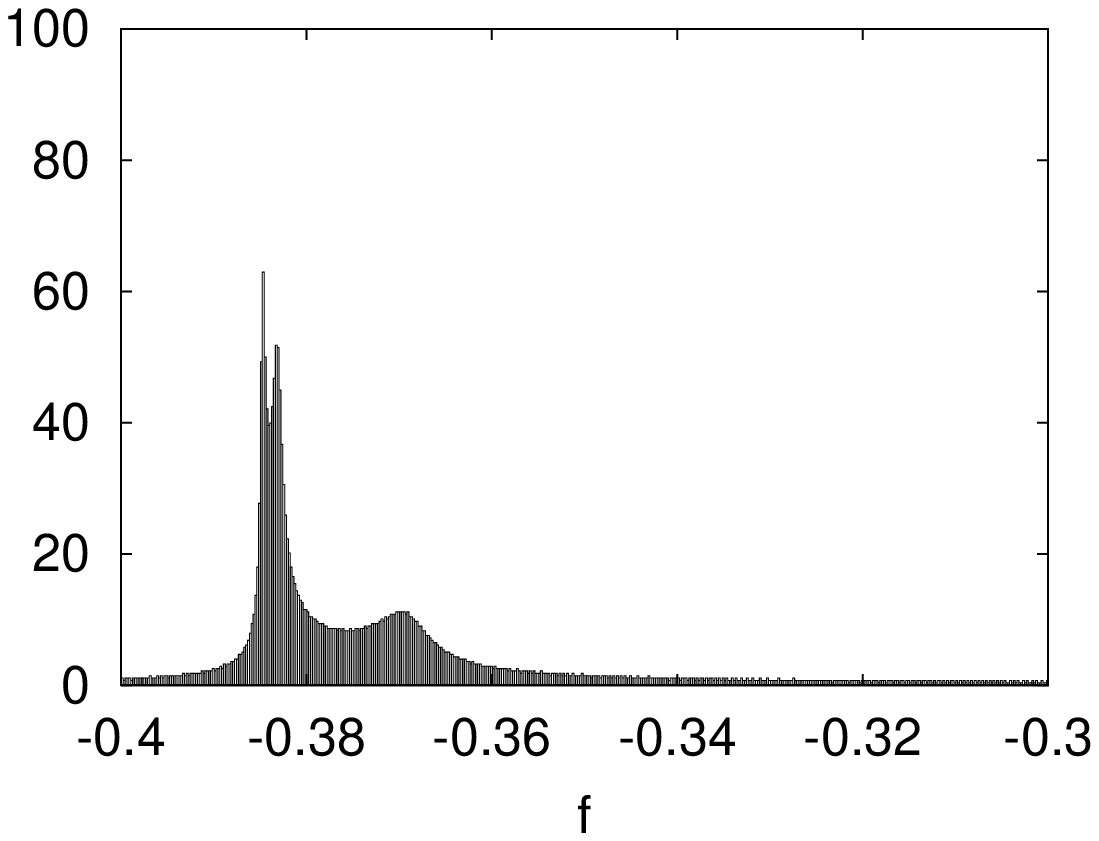,width=10cm} 
    \caption{
The histogram of the truncated
free energy density $f$ for $D=4$ at order 6.
The bin size is chosen to be very small (0.0002).
}
    \label{fig:freeE_histo_4d_sbin}
}

\FIGURE{\epsfig{file=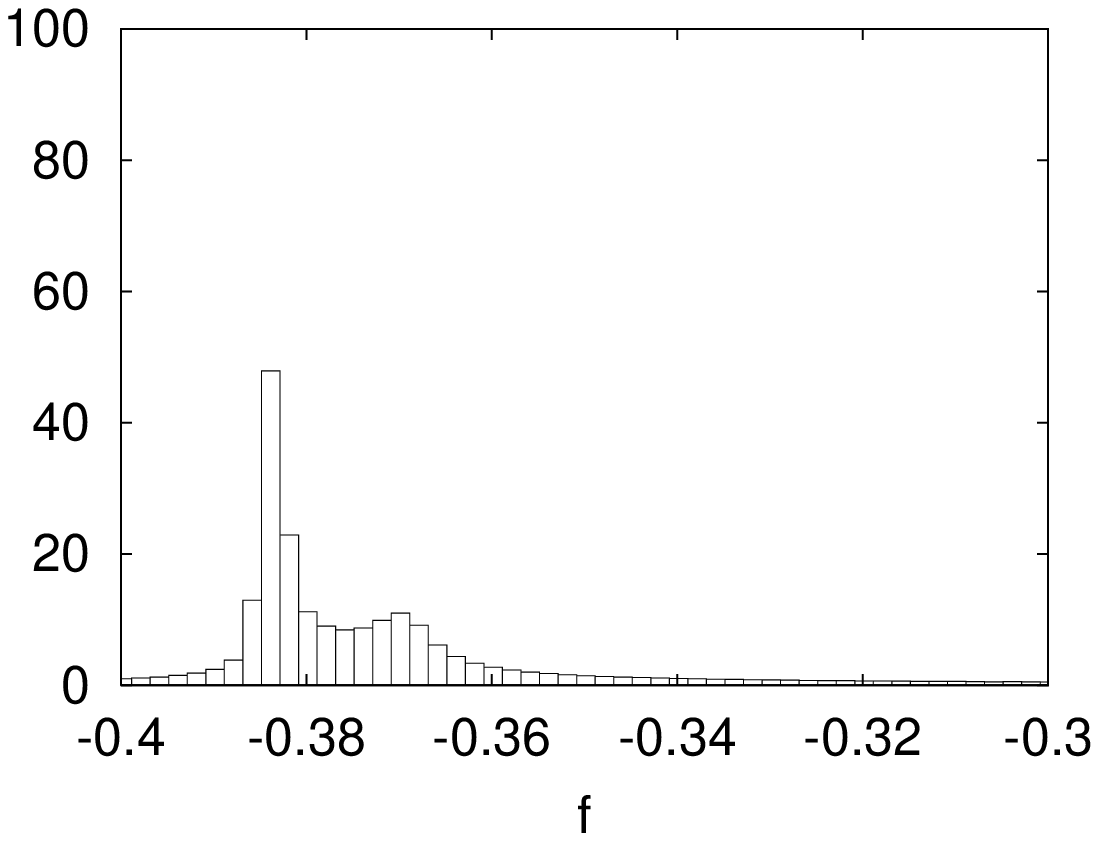,width=10cm} 
    \caption{
The histogram of the truncated
free energy density $f$ for $D=4$ at order 6.
The bin size (0.002) is chosen such that the highest bin is more than
twice as high as the second highest one.
}
    \label{fig:freeE_histo_4d_lbin}
}

\FIGURE{\epsfig{file=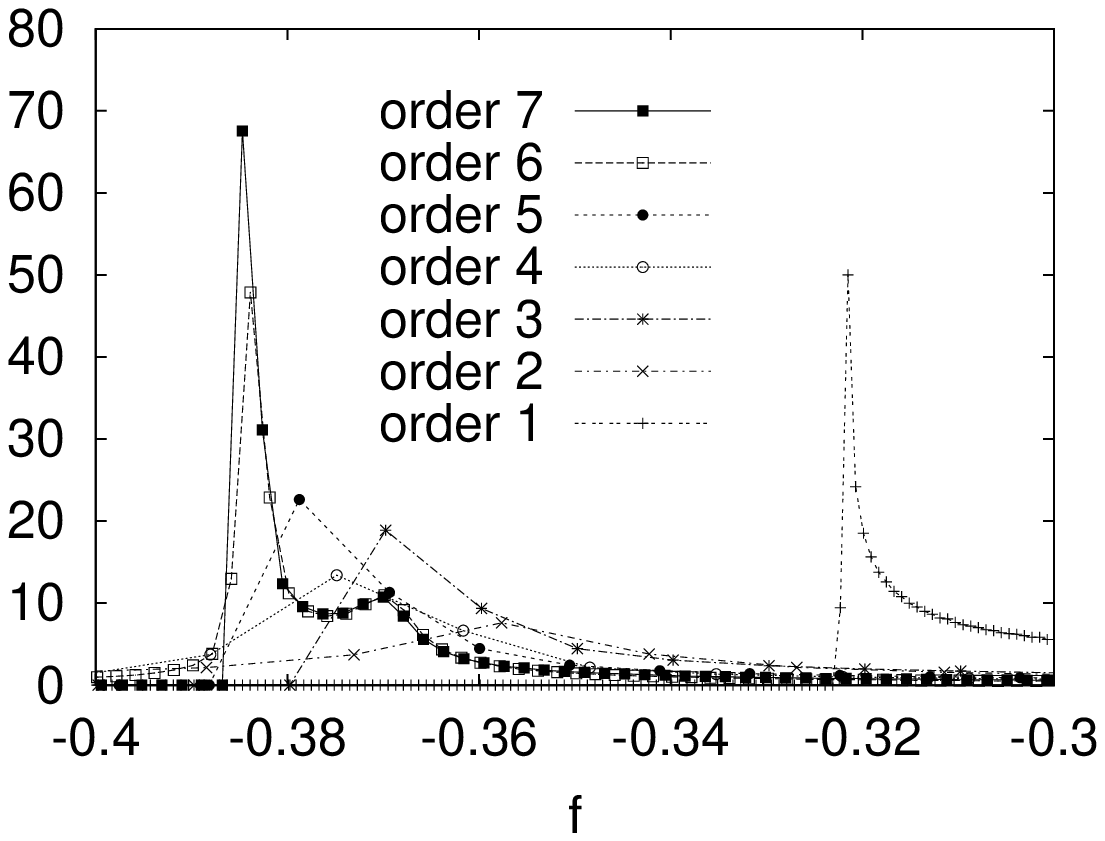,width=10cm} 
    \caption{
The histogram of the truncated free energy density $f$ for $D = 4$ 
at order 1, 2, $\cdots$, 7. 
The bin size is chosen carefully as described in the text at each
order.
The growth of the peak is clearly observed.
}
    \label{fig:freeE_histo_4d}
}


\end{document}